\documentclass[aps,pra,secnumarabic,
twocolumn,groupedaddress,showpacs,amsmath,amssymb]{revtex4}

\usepackage{bm}

\begin{document}
\title{GROUND STATE OF THE DOUBLE-WELL CONDENSATES
INTERACTING WITH TRAP OSCILLATIONS}
\author{L.~A.~Manakova}
\email[]{manakova@kurm.polyn.kiae.su} \affiliation{ RRC Kurchatov
Institute, Kurchatov Sq.\ 1, 123182 Moscow, Russia}

\begin{abstract}
In the present paper it is shown that the interaction between
classical anharmonic oscillations of a double-well trapped
condensate and excited Josephson states corresponding to a large
enough initial disbalance of the particle number generates their
bound state. The bound state can realize an absolute minimum of
the thermodynamic energy. The transition to new ground state is a
second-order phase transition. The existence of the bound state
implies that the Josephson states can be detected by observing the
change in the condensate shape.
\end{abstract}

\maketitle

\section{Introduction}

The realization of two-component condensates \cite{My} as well as
condensates in a double-well potential \cite{Science} has
attracted considerable interest, both experimental and theoretical
(see, e.g., \cite{Leg} and references therein) to the phenomena
caused by phase coherence of two condensate modes. In the work
\cite{Science} spatial quantum coherence was observed by means of
interference pattern in two overlapping condensates. This
interference pattern was conformed in the work \cite{Rohrl} with
the numerical simulation of the Gross-Pitaevskii equation. In the
work \cite{HM} coherent oscillations of the relative populations
have been observed in driven two-component condensates with
different internal states. As is well known, a bright
manifestation of phase coherence is Josephson effect. In numerous
studies devoted to the Josephson effect in systems of two
condensates in different internal states \cite{HM}, \cite{WW} or
in a double-well potential \cite{ZSL}-\cite{Walls} coherent
Josephson oscillations are considered for various dynamical
regimes caused by the competition between tunneling and
intracondensate interaction (nonlinearity). In the work \cite{ZSL}
the Josephson coupling energy is calculated for small-amplitude
oscillations in a double-well potential. There are estimated
damping effects due to the normal currents at finite temperature.
In the work \cite{Walls} it is shown that for a relatively weak
interaction, the particle number oscillations between the
condensates are complete. They are suppressed when the total
number of atoms in the condensates exceeds a critical value and
the behavior of the system is governed by the nonlinearity.
Nonlinear Josephson-type oscillations in the relative oscillations
of driven two-component condensates are studied in the work
\cite{WW}. Decoherence effects and quantum corrections to
mean-field solutions have been considered in \cite{You1},
\cite{You}. In \cite{MZ} the damping effects (even at zero
temperature) of the Josephson current are derived  by means of an
functional integral approach. The detailed treatment of the
nonlinear classical dynamic of the condensates in a double-well
potential has been performed in the works \cite{RS},\cite{RS1}. In
the work \cite{PS} the quantum and thermal fluctuations of the
phase are studied for the condensates in the double well
potential.

It should be emphasized that experimental observation of the
Josephson effect is difficult because the small energy splitting
associated with Josephson coupling means that thermal and quantum
fluctuations will destroy the phase coherence between two
condensates even at the lowest achievable temperatures \cite{MZ},
\cite{PS}. While the energy splitting can be increased, for
instance, by lowering of the barrier height, it then becomes
comparable with that of motion states of the condensates.

However, the problem of the interaction between the Josephson
degrees of freedom and states of motion (oscillations) of the
trapped condensate has yet to be analyzed. The present paper
focuses on a mechanism of formation of the bound state of the
Josephson degrees of freedom and trap oscillations due to their
interaction. The mechanism proposed in what follows may be
important for detection of the excited Josephson states.

The results achieved in this paper may be listed as follows.

1. To formulate the problem of the interaction between the
Josephson and oscillation degrees of freedom in an adequate
manner, the quantizied spectrum of the particle number generated
by Josephson coupling is derived. The states of this spectrum
represent the quantum analogue of the nonlinear coherent Josephson
oscillations considered in \cite{RS}. In what follows, the states
of the quantizied spectrum are named as Josephson states. The
spectrum is highly nonequidistant and has the logarithmic
singularity in the density of states at the $2E_J$ energy, where
$E_J$ is the Josephson coupling energy. As is shown in what
follows, any Josephson state can be realized by means of a given
initial disbalance of the particle number in two condensates.

2. There is considered the interaction between the trap
oscillations and excited Josephson states corresponding to the
large enough initial disbalance of the particle number. It is
shown that this interaction is responsible for the formation of
the bound state of $\bar{n}_m\gg 1$ oscillation quanta with the
Josephson state corresponding to the definite initial disbalance.
In the Thomas-Fermi approximation at $\mu\gg \omega_0$, where
$\mu$ is the chemical potential of the condensates, $\omega_0$ is
the characteristic frequency of a trap, $h=1$, the bound state
arises in the region of the dense oscillation spectrum. In this
region the level separations are small in comparison with the
harmonic oscillation frequency $\omega_0$. The equilibrium  values
of the oscillation quanta and initial disbalance of the particle
number are coupled self-consistently and can realize an absolute
minimum of the thermodynamic energy at the large enough
interaction. The thermodynamic average $\bar{n}_m\neq 0$ generates
the equilibrium distortion of the condensate shape. It allows us
to detect the Josephson states by observing the change in the
condensate shape.

\section {The quantum spectrum of the particle number}

A Josephson coupling is realized for the condensates in an
symmetrical/asymmetrical double-well potential formed by two
different traps with a barrier between them \cite{Science}. The
barrier is created by the laser light, and its height is directly
proportional to the laser power and, thus, can be varied with
ease. The proposed mechanism is also suitable for the condensates
in different internal states in the same trap. Experimentally,
this may be a superposition of two ${\rm Rb}^{87}$ condensates in
the $|F=1,m_F=-1\rangle, |F=2,m_F=1\rangle$ states \cite{HM},
\cite{WW}. In the case of a weak Josephson coupling, the basis
states are the self-consistent ground states in two condensates
separately. The wave function of the condensate with Josephson
coupling takes the form of a superposition of these states,
namely, $\Psi({\bf r},t)=\psi_1({\bf r})a_1(t)+\psi_2({\bf
r})a_2(t)$, where $\psi_i({\bf r})$ are the normalized solutions
of the Gross-Pitaevskii equation,
$a_i(t)=N_i^{1/2}(t)e^{i\theta_i(t)}$, $i=1,2$, $N_{1,2}$,
$\theta_i(t)$ are the particle numbers and phases of each
condensate.

As is well known \cite{Leg}, \cite{ZSL}, \cite{Walls}, \cite{RS1}
the Hamiltonian of two condensates with Josephson coupling has the
form
\begin{equation}
H_{J}-E_0=E_C(\Delta N)^2-2E_J\cos\phi,
\label{eq:14}
\end{equation}
here $\Delta N=N_1-N_2$, $E_C=\partial \mu/\partial N$, $\mu\equiv
\mu_1=\mu_2$ are the chemical potentials, $N$ are the total
particle numbers, $\phi=\theta_1-\theta_2$ is the relative phase
of the condensates. The quantities $E_C$, $E_J$ depend on the
total particle number $N$. $(\Delta N)$, $\phi$ are canonically
conjugate variables. In (\ref{eq:14}) the energy origin is the
mean-field summary energy of the condensates, namely, $E_0\equiv
\mu N$.

The quantization of the Hamiltonian (\ref{eq:14}) produces the
spectrum of the particle number in the Josephson potential
$E_J\cos\phi$. As is shown in what follows, any Josephson state
can be realized by means of a given initial disbalance of the
particle number. For this reason, it is interesting to obtain the
complete spectrum generated by the Hamiltonian of (\ref{eq:14}),
and to represent it as a function of the initial disbalance.

The Schrodinger equation for the Hamiltonian (\ref{eq:14}) is
derived by means of the quantization rule: $(\Delta N)\rightarrow
-i\partial/\partial\phi$. As a result, we obtain Mathieu's
equation
\begin{equation}\label{eq:15}
\left[-\lambda\frac{d^2}{d\phi^2}-2E_{J}\cos\phi_r\right]\Psi
=\varepsilon\Psi,\;\;\; \varepsilon\equiv H_J-E_0.
\end{equation}
For $\varepsilon>2E_{J}$ this equation has a continuous spectrum.
The states of this spectrum correspond to the classical states
with unlimited phase change: $-\infty<\phi<+\infty$ that are named
as the self-trapping states in the works \cite{RS,RS1}. In the
region $-2E_{J}<\varepsilon<2E_{J}$ Eq.(\ref{eq:15}) has the
discrete spectrum. It corresponds to the region of finite motion
of  Hamiltonian (\ref{eq:14}), where the relative phase changes
within $-\arccos(\varepsilon/2E_{J}) < \phi <
\arccos(\varepsilon/2E_{J})$ for every $\varepsilon$. In the
Josephson regime at $E_J\gg E_C$ \cite{Leg}, the number of levels
in a well is large and the discrete spectrum is specified by the
Bohr-Sommerfeld formula:
\begin{equation}
\begin{split}
 &\quad\nu(\varepsilon_{\nu})=\oint\frac{d\phi_r}{\pi}\Delta
 N(\phi_r;\varepsilon_{\nu}) \\
 =&\oint\frac{d\phi_r}{\pi}\Big[\frac{1}{E_C}
 \left(\varepsilon_{\nu}+2E_{J}\cos\phi\right)\Big]^{1/2} \\
 &=\nu_c\Big[E(\kappa)-(1-\kappa^2)K(\kappa)\Big];
\end{split}
\label{eq:16}
\end{equation}
\vspace*{-0.3cm}
\begin{equation}\nonumber
 \nu_c=\frac{8}{\pi}\left(\frac{E_J}{E_C}\right)^{\!\!1/2}\!\!;\;\;\;
 \kappa^2=\frac{\varepsilon_{\nu}+2E_J}{4E_{J}};\;\;\;\nu_c\gg 1;
\end{equation}
$K(\kappa)$ and $E(\kappa)$ are complete elliptic integrals of the
first and second kinds. The density of states
$\rho_d(\varepsilon_{\nu})$ results from Eq.(\ref{eq:17}) and is
equal to
\begin{equation}
\rho_d(\varepsilon_{\nu})=\frac{1}{2\pi}\frac{d\nu}{d\varepsilon_{\nu}}=
\frac{K(\kappa)}{\pi^2\omega_m}.
\label{eq:18}
\end{equation}
The level separations in (\ref{eq:16}) are given by the following
expression
\begin{equation}
\omega(\varepsilon_{\nu})=\frac{d\varepsilon}{d\nu}=\frac{\pi\omega_m}{2K(\kappa)},\;\;
\omega_m=2(E_C E_{J})^{1/2}.
\label{eq:17}
\end{equation}
At $(E_{J}/E_C)^{1/2}\gg 1$ we have the relation $\omega_m\ll
2E_{J}$. The quantity $\omega_m$ determines the maximum splitting
of the levels in the Josephson well. In what follows, the states
with $\nu<\nu_c$ are denoted as "libration" states.

In the region of $\nu\geq \nu_c$ the $\nu(\varepsilon)$ dependence
and density of states are determined by the following expressions
\begin{equation}
\nu(\varepsilon)=\frac{4}{\pi^2}
\left(\frac{E_{J}}{E_C}\right)^{1/2}\kappa E(\kappa^{-1});\;\;\;
\rho_c(\varepsilon)=
\frac{1}{2\pi^2}\cdot \frac{\kappa^{-1}K(\kappa^{-1})}{\omega_m}.
\label{eq:19}
\end{equation}
The states (\ref{eq:19}) with $\nu>\nu_c$ are named as the
self-trapping states.

Eqs.(\ref{eq:16}), (\ref{eq:19}) imply that
\begin{equation}
\begin{split}
&\varepsilon(\nu)\approx -2E_J+\omega_m \nu;\;\;\; 1\ll \nu\ll
\nu_c,\\
&\varepsilon(\nu)\approx 4\pi^2 E_C\nu^2;\;\;\; \nu\gg
\nu_c.
\end{split}
\label{eq:21}
\end{equation}
At the same time, it is easy to show that $d^2\varepsilon(\nu)/d\nu^2<0$ at
$\nu<\nu_c$ and $d^2\varepsilon(\nu)/d\nu^2>0$ at $\nu>\nu_c$.
At $\nu=\nu_c$ the curve $\varepsilon(\nu)$ has a flex point.

Since the energy is conserved, the state with a given $\nu$ value can be
realized by means of the definition of the initial values of $(\Delta N)_0$
and $\phi(0)$. Namely,
$\varepsilon(\nu)=E_C(\Delta N)_0^2-2E_J\cos\phi(0)$.
Supposing that $\phi(0)=0$, we obtain the following relation between $\nu$ and
$(\Delta N)_0$
\begin{equation}
\varepsilon(\nu)=-2E_J+E_C(\Delta N)_0^2
\label{eq:22}
\end{equation}
Using Eqs.(\ref{eq:21}), (\ref{eq:22}), we arrive at the
expressions
\begin{equation}
\begin{split}
&\nu=\frac{1}{2\pi}|(\Delta N)_0|;\;\;\;\text{for the
self-trapping (sf) states,}\\
&\nu=\left(\frac{E_C}{\omega_m}\right)(\Delta N)_0^2;\;\;\;\text{
for the ``libration'' (l) states.} \label{eq:202}
\end{split}
\end{equation}

Combining (\ref{eq:18}), (\ref{eq:19}), we come to the depedence
\begin{equation}
\rho_{d,c}(\varepsilon)\sim \omega_m^{-1}\ln|1-\varepsilon/2E_{J}|^{-1};\;\;\;
\varepsilon\rightarrow 2E_{J}^{\pm}
\label{eq:20}
\end{equation}
Thus, there appears a new logarithmic
singularity at the boundary separating the libration and self-trapping spectra.

\section {The interaction  of the Josephson states and condensate oscillations}

In this and next sections it is shown that the spectrum of the
system can change drastically due to the interaction between the
excited Josephson states (\ref{eq:21})-(\ref{eq:202}) with large
enough $\nu$ values and oscillations of the condensate.

The interaction can be realized by the following mechanisms.
First, the interaction is secured if we allow for the dependence
of $E_C$ in Eq.(\ref{eq:22}) on the atom displacements. The latter
are generated by the condensate oscillation. Second, the
interaction can be realized by applying two-photon traveling-wave
laser pulse with the Rabi frequency $\Omega$. The pulse both
creates the condensates with the different particle numbers and
induces the interaction of atom's displacements with the excited
Josephson states corresponding to the particle number disbalance
created by the pulse. General description proposed in what follows
is independent on the specific mechanism producing the
interaction.

Let us consider the classical states of motion of the condensate.
These states may be described in terms of the complex amplitudes
$a^*,a=n^{1/2}e^{\pm i\varphi_1}$, where $n=\langle
a|\hat{a}^+\hat{a}|a \rangle=|a|^2$ is the average number of
quanta in the coherent state $|a \rangle$. The variables
$n,\varphi_1$ are canonical. By classical state of motion we mean
that it's number of quanta is very large, $n\gg 1$. It is
convenient to specify the relation between the $a, a^*$ amplitudes
and $\hat{a},\hat{a}^+$ operators  by the following way:
$a=N^{-1/2}\hat{a}$. At this, the commutator of $a,a^*$ is equal
to zero with macroscopic accuracy: $[a,a^*]=1/N\rightarrow 0$. The
Hamiltonian of the motion states can be written in the form:
$N\epsilon(n)$.

For the quasiclassical Josephson states with $\nu\gg 1$, the
$c_{\nu},c_{\nu}^*$ amplitudes may be written in the form
$c_{\nu}=\nu^{1/2}e^{i\varphi_2}$. However, it is convenient to
rewrite $\varepsilon(\nu)$, $c_{\nu}$ in terms of the variable
$x=|(\Delta N)_0|/N^{1/2}\gg 1$. Combining this inequality with
the requirement that $x=|(\Delta N)_0|/N\ll 1$ we arrive at the
conditions for the $x$ values:
\begin{equation}
1\ll x\ll N^{1/2}.
\label{eq:020}
\end{equation}
Using Eqs.(\ref{eq:21}),(\ref{eq:22}), we find that $\nu=\nu(x)$,
$\varepsilon(\nu)=N(-E_J/N + E_C x^2)$. In general case, the
$\nu(x)$ dependence is implicit. It is determined by
Eqs.(\ref{eq:16}), (\ref{eq:19}), (\ref{eq:22}). However, in the
particular cases of the "libration" ($\varepsilon(\nu)\ll E_J$)
and self-trapping ($\varepsilon(\nu)\gg E_J$) states, the
relations between $\nu$ and $(\Delta N)_0$ can be represented in a
simple form, as is seen from (\ref{eq:202}). By means of
Eqs.(\ref{eq:22}), (\ref{eq:202}), we come to the following
expressions
\begin{equation}
\begin{split}
H_0\equiv N&\varepsilon_0(n,x)=N[-\frac{E_J}{N}+\epsilon(n)+E_C
x^2]\\
&=N[-\frac{E_J}{N}+\epsilon(n)+\varepsilon_J(x)],
\end{split}
\label{eq:8}
\end{equation}

\begin{equation}
\begin{split}
&\quad c^{(sf)}_{\nu}=N^{1/4}x^{1/2}e^{i\varphi_2}\equiv
N^{1/4}c_x^{(sf)};\\
&c^{(l)}_{\nu}=N\left(\frac{E_C}{\omega_m}\right)^{\!\!1/2}\!\!xe^{i\varphi_2}\equiv
N\left(\frac{E_C}{\omega_m}\right)^{\!\!1/2}\!\!c_x^{(l)}.
\end{split}
\label{eq:6}
\end{equation}

For any mechanism producing the interaction between two
subsystems, it can be written in the form of a multiple Fourier
series in $\varphi_1$, $\varphi_2$

\begin{equation}
\begin{split}
&\quad
H_{int}=N\sum_{k_1k_2}[g_{k_1k_2}(N)a^{*k_1}c^{k_2}_{\nu}+c.c.]\\
& =N\sum_{k_1,k_2}[g^{(sf,l)}_{k_1k_2}(N)n^{k_1/2}x^{\alpha
k_2}e^{i(k_1\varphi_1- k_2\varphi_2)}+c.c.].
\end{split}
\label{eq:a7}
\end{equation}
Here $\alpha_{sf}=1/2, \alpha_{l}=1$, $k_1,k_2$ are integer. For
the sake of simplicity, we disregard the phase-independent
interaction. Using Eq.(\ref{eq:6}), we obtain
\begin{equation}
g_{k_1k_2}^{(sf)}(N)= g N^{-1+k_2/4};\;\;\;
g_{k_1k_2}^{(l)}(N)=g N^{-1+k_2}\cdot
\left(\frac{E_C}{\omega_m}\right)^{k_2/2}.
\label{eq:7a}
\end{equation}
The constant $g$ is specified by the concrete mechanism producing the interaction.
Let us assume that the term with the phase $\phi_k^r=k_{1r}\varphi_1-k_{2r}\varphi_2$, which varies anomalously slowly with time,
can be set off in sum (\ref{eq:a7}).
It is possible to make under two conditions
\begin{equation}
\begin{split}
k_{1r}\left(\frac{d\epsilon(n)}{dn}\right)=k_{2r}&\left(\frac{d\varepsilon_J(x)}{dx}\right),\;\;\;
\epsilon'_n\equiv \left(\frac{d\epsilon(n)}{dn}\right),\\
&x_m=\frac{k_{1r}\varepsilon'_n}{2k_{2r} E_C}.
\end{split}
\label{eq:9}
\end{equation}

\begin{equation}
\left(\frac{d^2 H_0}{dx^2}\right)_{x=x_m} (\Delta x)_{max}\gg
\left(\frac{\partial H_{int}}{\partial x}\right)_{x=x_m}.
\label{eq:9a}
\end{equation}
Eq.(\ref{eq:9a}) is written with taking into account
that $H_0$, $H_{int}$
are functions of a single dynamic variable, e.g., x.
The $\epsilon'_n$ quantity defines the level
separations of an oscillation spectrum.

As is shown in what follows, the condition (\ref{eq:9}) is
equivalent to that of the minimum of the function
$\varepsilon_0(n,x)$ over x. In turn, when the minimum exists, it
can provide the principal contribution into the thermodynamic
functions.

In addition,  the condition (\ref{eq:9}) implies that the phase
$\phi_k^r$ is an approximate integral of motion in the absence of
changing $x$ near $x_m$: $d\phi_k^r/dt\approx k_{1r}\partial
H_0/\partial n-k_{2r}\partial H_0/\partial x\approx 0$.

The inequality  (\ref{eq:9a}) implies that the width of the
near-minimum region is large at the characteristic interaction
variation scale. From Eqs.(\ref{eq:9}), (\ref{eq:9a}) one can
obtain that time changing the $\phi_k^r$ phase is proportional to
$(d^2 H_0/dx^2)_m \Delta x$, where $\Delta x$ is the change in $x$
near the $x_m$ value. The maximum value $(\Delta x)_{max}$
specifies the width of the near-minimum region in that
$d\phi_k^r/dt \sim \Delta x$. The estimation for $(\Delta
x)_{max}$ is given in what follows. Thus, the leading term in sum
(\ref{eq:a7}) has the form
\begin{equation}
H_{int}^{(r)}=Ng_k^{(sf,l)}(N;n,x)\cos\phi_k,\;\;\;\phi_k=k_{1r}\varphi_1-k_{2r}\varphi_2;
\label{eq:7}
\end{equation}
All remaining terms in this sum are rapidly oscillating
perturbations and will be disregarded in this work. Here and below
the index $k$ in $g_k$ and $\phi_k$ denotes a set of
$k_{1r},k_{2r}$. One can easy to show that aside from the energy
$H=N[\varepsilon_0(n,x)+g_k(N;n,x)\cos\phi_k)]$ the system in
question has an additional integral of motion:
$n_0=n/k_{1r}+x/k_{2r}$, $dn_0/dt=0$. Owing to this, the condition
(\ref{eq:9}) is equivalent to that of the minimum of
$\varepsilon_0(n_0,x)$  over x at a given $n_0$ value, as is
mentioned above. Using Eqs.(\ref{eq:9})-(\ref{eq:7}), it is
straightforward to write the Hamiltonian $H_m=H_0+H_{int}^{(r)}$
near the minimum to the first nonvanishing order in $\Delta x$:
\begin{equation}
H_{m}=N\left[\varepsilon_{0}(n_m;x_{m})+\left(\frac{d^2\varepsilon_{0}}{dx^2}\right)_m
(\Delta x)^2-g_{km}\cos\phi_k\right],
\label{eq:10}
\end{equation}
$n_m=n_0-k_{1r}x_m/k_{2r}$, $(d^2\varepsilon_{0}/dx^2)_m=2E_c$,
$g_{km}=g_k^{(sf,l)}(N;n_m,x_m)$. Terms with the derivatives of
$H_{int}$ are absent in Eq.(\ref{eq:10}) due to the condition
(\ref{eq:9a}).

Using that $E_C\sim \omega_0(a/a_0)^{2/5}N^{-3/5}$ in the
Thomas-Fermi approximation \cite{Pit} (here $a$, $a_0$ are the
scattering and oscillator lengths, respectively), we can represent
the range of $1\ll x_m \ll N^{1/2}$ in the form
\begin{equation}
\frac{1}{N}\left(\frac{Na}{a_0}\right)^{2/5}\ll \frac{k_{1r}}{k_{2r}}
\left(\frac{\epsilon_n^{'}}{\omega_0}\right)\ll
\frac{1}{N^{1/2}}\left(\frac{Na}{a_0}\right)^{2/5}.
\label{eq:10a}
\end{equation}
As is known \cite{Pit}, the relation $(Na/a_0)\gg 1$ occurs  in
the Thomas-Fermi approximation. However,
$N^{-1/2}(Na/a_0)^{2/5}\ll 1$. Owing to this relation, the
condition (\ref{eq:10a}) (or, what is the same, the condition
(\ref{eq:9})) specifies the region of the dense oscillation
spectrum, where $\epsilon_n^{'}\ll \omega_0$. Both here and in
what follows, we suppose that $k_{1r}=k_{2r}=1$ for the sake of
simplicity.

From Eq.(\ref{eq:10}) the $(\Delta x)_{max}$ can be estimate to be
$(\Delta x)_{max}\sim (g_{km}/E_C)^{1/2}$. Hence, the condition
(\ref{eq:9a}) takes the form:
\begin{equation}
\frac{|x_{m}-n_{m}|}{x_{m}n_{m}}\ll\left(\frac{E_C}{g_{km}(N)}\right)^{1/2},
\label{eq:001}
\end{equation}
In what follows (see Eq.(\ref{eq:33})), it will be shown that the
relation (\ref{eq:001}) is fulfilled with macroscopic accuracy.

\section {The ground state}

At a fixed $n_0$ value the principal contribution to the partition
function comes from the neighbourhood of the minimum at $x=x_m$.
The expression for $Z(n_0;x_{m};T)$ is equal to
\begin{equation}
\begin{split}
&Z(n_0;x_{m};T)=const\int\limits^{\infty}_{-\infty} d(\Delta x)
\int\limits^{\pi}_{-\pi} d\phi_k e^{-\beta H_m(n_0,(\Delta
x),\phi_k)}\\ &=\frac{const}{(\beta N E_C)^{1/2}} \exp
\left[-\beta N \varepsilon_{0}(N;n_m;x_{m})+\ln I_0(\beta N
g_{km})\right],
\end{split}
\label{eq:11}
\end{equation}
where $\beta=1/T$, $T$ is temperature, $I_0(x)$ is the modified
Bessel function. Eq.(\ref{eq:11}) implies that the expression for
the free energy of the system has the form
\begin{equation}
F=N\varepsilon_{0}(n_{m};x_{m})+\frac{1}{2}T\ln(\beta N E_c)
-T \ln I_0(\beta N g_{km}),
\label{eq:13}
\end{equation}

Using Eq.(\ref{eq:13}), we come to the following equation for the $\bar{n}_m$
value realizing the minimum of the free energy
\begin{equation}
\left(\frac{d \varepsilon_{0m}}{d n_m}\right)_{n_m=\bar{n}_m}=
\left(\frac{d g_{km}}{d n_m}\right)_{n_m=\bar{n}_m}\cdot
\frac{I_1(\beta N g_{km})} {I_0(\beta N g_{km})}, \label{eq:14a}
\end{equation}
here $I_1(x)=I'_0(x)$. In addition to $\bar{n}_m$, the
thermodynamic average of $\cos\phi_k$ may be determined from
Eqs.(\ref{eq:11}) or (\ref{eq:13}). This average is equal to
\begin{equation}
\langle\cos\phi_k\rangle_T=-\frac{\partial \ln Z}{\partial(\beta N g_{km})}=
\frac{\partial F}{T\partial(\beta N g_{km})}=
\frac{I_1(\beta N g_{km})}{I_0(\beta N g_{km})},
\label{eq:16a}
\end{equation}
The order parameters $\bar{n}_m$, $\langle\cos\phi_k\rangle_T$
describe new coherent state. There is the bound state of the
$\bar{n}_m$ oscillation quanta and Josephson state generated by
the initial disbalance  of the particle number that corresponds to
the $x_m$ value. In addition, this state has the equilibrium phase
coherence factor $\langle\cos\phi_k\rangle_T$. The $\bar{n}_m\neq
0$ value provides the equilibrium distortion of the condensate
shape. The aboveobtained equations imply that the shape distortion
is self-consistently coupled to the $x_m$ value defining the
equilibrium initial disbalance of the particle number.

At $T=0$ the $\bar{n}_m$ value realizes the minimum of the
thermodynamic energy
\begin{equation}
E_m=N[\varepsilon_{0}(N;n_{m},x_{m})-g_{km}(N;n_{m},x_{m})].
\label{eq:17a}
\end{equation}
To determine $\bar{n}_m(T=0)$, it is suitable to use the following
consideration. As is well known, the level separations
$\epsilon'_n$ change slowly in dependence on $n$ within the range
of dense (quasiclassical) spectrum. For this reason, we can
suppose that $\epsilon_n^{'}\approx const\equiv \omega_b\ll
\omega_0$. On this assumption, the $x_m$ value does not depend on
$n_m$ and the $\bar{n}_m$ quantity is equal to
\begin{equation}
\bar{n}_{m}^{1/2}=\frac{g^{(sf,l)}(N)x_m^{\alpha}}{2\omega_b},\;\;
\alpha_{sf,l}=\frac{1}{2},1,
\label{eq:15a}
\end{equation}
(It is worth noting that $(\partial^2 E_m/\partial n_m^2)>0$).
Both here and in what follows, the denotation
$g_{k=1}^{(sf,l)}(N)\equiv g^{(sf,l)}(N)$ is used. Taking
Eq.(\ref{eq:15a}) into account, one readily gets
$$\bar{g}_{1m}^{(sf)}=\frac{g^{(sf)2}(N)x_m}{2\omega_b},\;\;\;
\bar{g}_{1m}^{(l)}=\frac{g^{(l)2}(N)x_m^2}{2\omega_b},$$
\begin{equation}
\begin{split}
&\frac{E_m^{(sf)}}{N}=-\frac{E_J}{N}-\frac{g^{(sf)2}(N)}{8E_C}
\left(1-\frac{8E^2_c x_m^2}{g^{(sf)2}(N)}\right),\\
&\frac{E_m^{(l)}}{N}=-\frac{E_J}{N}-\frac{g^{(l)2}(N)x_m}{16E_C}
\left(1-\frac{16E^2_c x_m}{g^{(l)2}(N)}\right).
\end{split}
\label{eq:20a}
\end{equation}

The expressions for energies imply, first, that we obtain the
minimum in the region of the dense enough oscillation spectrum,
which satisfies the condition (\ref{eq:10a}). The minimum
corresponds to the formation of the bound state for the
$\bar{n}_m, x_m$ values. Second, as is seen from
Eq.(\ref{eq:20a}), the absolute minimum of $E_m$ can be realized
within the ranges of
\begin{equation}
1\ll x_m < \frac{g^{(sf)}(N)}{8E_C};\;\;\;1\ll x_m < \frac{g^{(l)2}(N)}{16E^2_C}.
\label{eq:30}
\end{equation}
These conditions are met when the interaction matrix elaments
$g^{(sf,l)}(N)$ are large enough. Let us estimate the condensate
parameters that are required for existence of the absolute
minimum. In the Thomas-Fermi approximation, the inequality
$g^{(sf)}(N)\gg E_C$ occurs provided the total particle number is
not very large, namely,
\begin{equation}
N\ll \frac{g}{\omega_0}\left(\frac{a_0}{a}\right)^{8/3}.
\label{eq:31}
\end{equation}
In turn, the relation $g^{(l)2}(N)\gg E^2_C$ holds true within the
range of
\begin{equation}
N^{0.1}\frac{g^2}{\Omega^{1/2}\omega^{3/2}_0}\left(\frac{a_0}{a}\right)^{0.6}\gg
1, \label{eq:32}
\end{equation}
Here we use that $E_J=\Omega N$. The condition (\ref{eq:32}) is fulfilled for all
admissible parameters, if $g^2/(\Omega^{1/2}\omega^{3/2}_0)\sim 1$.
It should be emphasized that the right-side inequalities in Eq.(\ref{eq:30}) are much
stronger than the condition $x_m\ll N^{1/2}$.

The transition to the state with $\bar{n}_m\neq 0$,
$\langle\cos\phi_k\rangle_T\neq 0$ is second-order. By imposing
that $\bar{n}_m\rightarrow 0$ at the transition temperature, one
gets from Eq.(\ref{eq:14a}).
\begin{equation}
T^{(sf,l)}_c=\frac{g^{(sf,l)2}(N)x_m^{\alpha}}{\omega_b}\cdot N.
\label{eq:18a}
\end{equation}
The dependences of the transition temperatures (\ref{eq:18a}) on
the total particle number are given by the expressions
\begin{equation}
T_c^{(sf)}\sim N^{-0.2},\;\;\;T_c^{(l)}\sim N^{0.8}.
\label{eq:34}
\end{equation}
Thus, the transition temperature $T_c^{(sf)}$ has the macroscopic
smallness in comparison with the $T_c^{(l)}$ temperature. Along
with the conditions (\ref{eq:31}), (\ref{eq:32}), this fact
implies that the libration Josephson state forms the bound state
with the condensate oscillation rather than the self-trapping
state.

\section {Concluding remarks}

We have found that the interaction between the Josephson and
oscillation states results in new coherent ground state of the
double-well trapped condensate. There is the bound state of the
anharmonic condensate oscillation and excited Josephson state. The
latter is specified by the definite initial disbalance of the
particle number. The $x_m$ value defining the disbalance is
self-consistently coupled to the equilibrium number of the
oscillation quanta entering the bound state. Along with the
$\bar{n}_m$ order parameter, this state has the equilibrium phase
coherence factor $\langle\cos\phi_k\rangle_T$, where $\phi_k$ is
the relative phase of the Josephson and oscillation modes.

It should be emphasized that the bound state arises in the
neighbourhood of the minimum of $\varepsilon_0(n_0;x)$ over $x$,
where the conditions (\ref{eq:9}), (\ref{eq:9a}) are met. The
condition (\ref{eq:9}) specifies the type of the interaction
between the Josephson and oscillation degrees of freedom. In
addition, it imposes the definite restrictions on the spectrum of
the oscillation states which may effectively interact with the
Josephson degrees of freedom. Namely, the density of oscillation
states should be large enough in order to satisfy the conditions
(\ref{eq:10a}). For instance, in an asymmetrical double-well
potential the energy $\epsilon(n)$ of it's classical oscillation
states has three branches. Two branches have energies of
$\epsilon_{1,2}(n)\leq V_b$, here $V_b$ is the barrier height.
There are the maximums at $\epsilon_{1,2}(n_{max})=V_b$. The third
branch has both the energy $\epsilon_{3}(n)\geq V_b$ and the
minimum at $\epsilon_{3}(n_{min})=V_b$. Thus, the regions of the
dense spectrum exist in the neighbourhood of the extremums of the
functions $\epsilon_{i}(n)$ ($i=1,2,3$). In the other words, in
the neighbourhood of barrier top.

As is shown in section III, the condition (\ref{eq:9a}) can be
represent in the form (\ref{eq:001}). Substituting the
aboveobtained expressions for $\bar{n}_m$, $\bar{g}_{1m}$ into
(\ref{eq:001}), we find that it takes the form
\begin{equation}
|x_{m}-n_{m}|^{(sf,l)}\ll \left(\frac{g_{1m}^{(sf,l)}(N)}{E_C}\right)^{1/2}.
\label{eq:33}
\end{equation}
This inequality is satisfied with macroscopic accuracy.

It is worth noting that the conditions (\ref{eq:31}),
(\ref{eq:32}) of the absolute minimum of the thermodynamic energy
are experimentally controlled by means of either the interaction
matrix element $g$ or the particle number $N$.

The existence of the bound state generates the equilibrium
distortion of the condensate shape specified by the $\bar{n}_m$
value. This mechanism can provide the experimental detection of
the excited Josephson states. The latter can be observed by
changing the condensate shape. In addition, the phase transition
to the new ground state occurs at $T=T_c$, where transition
temperatures $T_c^{(sf,l)}$ are defined by Eq.(\ref{eq:18a}).

\begin{acknowledgments}
I am grateful to Yu.M.Kagan and L.A.Maksimov for helpful
discussions.

This work was supported by the Russian Foundation for Basic
Research.
\end{acknowledgments}

\bibliography{ground}

\end{document}